\documentclass[3p,a4paper,12pt]{elsarticle}

\usepackage{amssymb}

\begin{document}

\begin{frontmatter}

\title{Fast Approximation Algorithms for Art Gallery Problems in Simple Polygons}

\author[First]{Dae-Sung Jang}
\ead{noshadowings@kaist.ac.kr}
\author[First]{Sun-Il Kwon}
\ead{iddaga@kaist.ac.kr}

\address[First]{Korea Advanced Institute of Science and Technology, Daejeon, Korea}

\begin{abstract}
We present approximation algorithms with $O(n^3)$ processing time for the minimum vertex and edge guard problems in simple polygons. It is improved from previous $O(n^4)$ time algorithms of Ghosh. For simple polygon, there are $O(n^3)$ visibility regions, thus any approximation algorithm for the set covering problem with approximation ratio of $\log(n)$ can be used for the approximation of $n$ vertex and edge guard problems with $O(n^3)$ visibility sequence. We prove that the visibility of all points in simple polygons is guaranteed by covering $O(n^2)$ sinks from vertices and edges : It comes to $O(n^3)$ time bound.

\end{abstract}


\end{frontmatter}


\newtheorem{thm}{Theorem}[section]
\newtheorem{lem}[thm]{Lemma}
\newproof{pf}{Proof}

\section{Introduction}
The art gallery problem is to find the minimum set of points $G$ such that any point in a polygon $P$ is visible from some point in the set $G$. The points in $G$ are called \emph{guards}. Two points $x,y$ are mutually \emph{visible} if all the points of convex combination of $x,y$ are in the polygon. The art gallery problem is proved as a NP-hard problem first by O'Rourke and Supowit \cite{Oro83-1} in polygons with holes. Lee and Lin \cite{Lee86} showed minimum vertex, edge and point guard problem in simple polygon are also NP-hard. It is known that the number of guards needed is at most $\lceil n/3\rceil$ for stationary point guards in simple polygon \cite{Chv75}. There are many results about related research \cite{Avi81,Ede84,Kah83,Oro83-2}.

A vertex guard problem and an edge guard problems are restricted version of original art gallery problem called point guard problem. The \emph{vertex guard} problem(VG problem) is to find the minimum set $G_v$ of vertices of polygon such that any point in the polygon is visible from some vertices in $G_v$. \emph{Edge guards} problem(EG problem) is similar to the VG problem but all points in $P$ should be weakly visible from some edges in the minimum set of edges $G_e$. A point $z$ in $P$ is \emph{weakly visible} from an edge $e$ of $P$ if there exists a point $u$ on $e$ such that $z$ and $u$ are visible.
Ghosh \cite{Gho87} presents $O(n^{5}\log{n})$ time algorithm for VG problem with a approximation ratio of at most $O(\log{n})$ times the minimum number of VGs. Aggarwal, Ghosh, and Shyamasundar \cite{Agg88} present $O(n^{4}\log{n})$ time algorithm for covering a polygon by star-shaped pieces that is at most $O(\log{n})$ times the optimal number of star-shaped pieces. Each star-shaped pieces can be visible from one point in its \emph{kernel} thus the same number of guards can cover the original polygon. Efrat and Har-Peled \cite{Efr06} give randomized approximation algorithms for VG problem in simple polygons with $O(nc_{opt}^{2}\log^4{n})$ time complexity and $O(\log{c_{opt}})$ approximation ratio. In polygons with $h$ holes, they present $O(nhc_{opt}^{3}polylog n)$ time algorithm with $O(\log{n}\log(c_{opt}\log{n}))$ approximation ratio. $c_{opt}$ is optimal number of VGs in the problem, and it can be $O(n)$ in the worst case of input. Therefore, even in expected time, the time complexity is greater than $O(n^3)$ in the worst case.

Recently, Ghosh \cite{Gho10} present new approximation algorithm run in $O(n^4)$ time in simple polygons and $O(n^5)$ time in polygons with holes for VG and EG problem. The approximation ratio $O(\log{n})$ is the same as previous results since his algorithm first discretizes the polygon into convex components with respect to visibility of vertices of the polygon, and after that, solves the problem as the set-covering problem with greedy heuristic. Ghosh argue that this type of solving technique, transforming art gallery problem into set-covering problem after discretizing the entire polygon, is the only known one leading to efficient approximation algorithms in terms of worst case running times and approximation bounds. The convex components are similar convex elements concept to the visibility regions \cite{Bos91,Bos02}. Visibility regions are constructed from the windows of vertices thus its edges divide the interior of the polygon whether corresponding vertex is visible or not. On the other hand, convex components are from line segments passing through any two vertices of the polygon. Therefore, there are neighboring components that their sets of visible vertices of polygon are exactly the same and thus these division is redundant from the view of visibility. By the previous work of Bose \cite{Bos91}, the number of visibility regions is $O(n^3)$ in simple polygon. Ghosh recognized that the same principles from the lemmas of \cite{Bos91} are also valid in convex components. $O(n^4)$ time complexity in simple polygons naturally comes from $O(n^3)$ number of elements, i.e. convex components, and $n$ sets each of which is for a vertex of polygon in set-covering problem.

We focus on the other property of visibility regions that Bose \cite{Bos91} also discovered. The number of sinks in the simple polygon is $O(n^2)$. In this report, we present new faster approximation algorithms for VG and EG problems than previous $O(n^4)$ time algorithm in simple polygon. It is based on the theorem that the number of sinks is much smaller than the number of whole visibility regions. From the next part, we define some terminologies and prove lemmas which conclude that the algorithms leads to $O(n^3)$ time complexity and $O(\log{n})$ approximation ratio.

\section{Approximation Algorithm for Vertex Guards}

A \emph{visibility polygon} $VP(P,z)$ denote the set of all points of $P$ that are visible from a point $z$ in $P$. For following terminologies, we use the same definitions in \cite{Bos91,Bos02}. A maximally connected subset $R$ of $P$ is a \emph{visibility region} if any two points in $R$ are visible from the same subset of vertices of $P$. The \emph{visibility set} $M_R$ of $R$ is the subset of vertices of $P$ visible from $R$. Two visibility regions are \emph{neighboring} if they share a common edge of their boundary. A \emph{sink} $s$ is a visibility region that has a visibility set $M_s$ such that $M_s \subset M_q$ for every neighboring visibility regions $q$ of it. A \emph{window} of a point $x$ denote a segment that is a part of an edge of $VP(P,x)$ and not contained in the boundary of $P$. The vertex of a window closest to $x$ is a \emph{base} and the other is an \emph{end}.

Sink has only incoming edges in the dual graph of the visibility region division. The \emph{dual graph} of visibility regions is an directed acyclic graph that connects each neighboring visibility regions that share a common edge and its direction towards the visibility region with the smaller visibility set. By one of lemmas and a corollary in \cite{Bos91}, Two visibility regions that share a common edge have the same visibility set except for one vertex and through the direction of corresponding edge of dual graph, there is a loss of visibility from one vertex. Let a VG-sink denote a sink of visibility regions for VG problem

\begin{figure}
\centerline{
    \includegraphics[width=6in]{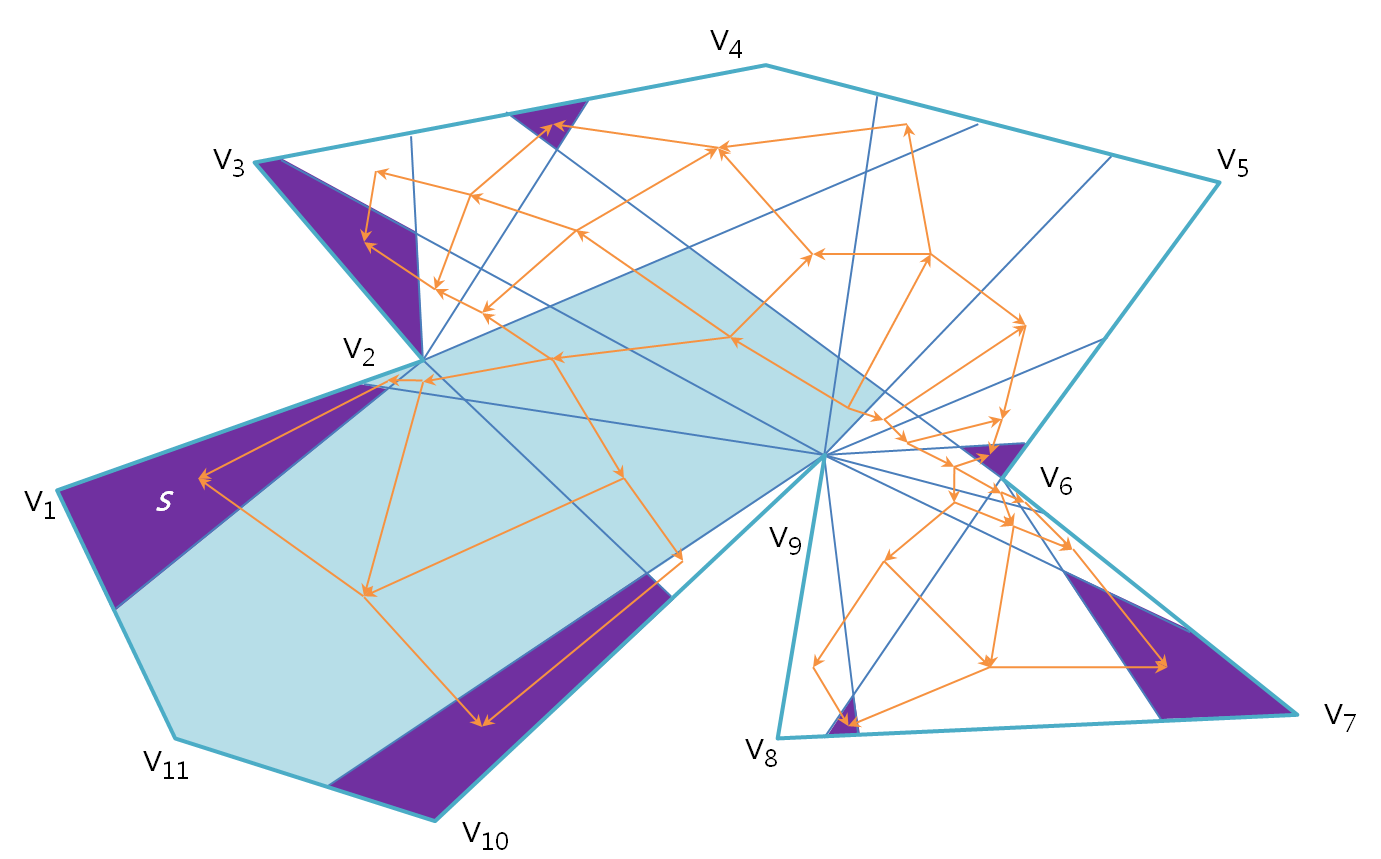}}
    \caption{Decomposition of a polygon into visibility regions and its dual graph. Shaded(sky blue) regions are elements of span $S$ of sink $s$ \label{fig1}}
\end{figure}

\begin{lem}\label{lemma1}
The set of VGs $G_v$ is the optimal solution of VG problem in a simple polygon $P$ if and only if $G_v$ is the minimum set that covers all VG-sinks.
\end{lem}
\begin{pf}
As it is noted before, in the dual graph of planar subdivision into visibility regions, a sink $s$ has only incoming edges. Since the dual graph is directed acyclic graph, there is a path to a sink from any node of dual graph. Suppose a region that $s$ is reachable from it and let the \emph{span} $K$ of sink $s$ denote the set of all such regions. Then the union of spans of all sinks are the same as $P$. Since, in a directed edge $(i,j)$ of dual graph, the visibility set of region $i$ includes the visibility set of region $j$, a sink has the set of visible points which are also visible from any element in its span $K$. Therefore, if $G_v$ is the minimum set that covers all VG-sinks, it covers all visibility regions; any point in $P$ is visible from some point in $G_v$. To assure $G_v$ is the optimal solution of VG problem when $G_v$ is the minimum set that covers all VG-sinks, let suppose there is a set $H_v$ the optimal solution of VG problem that has smaller cardinality than $G_v$. Since any point in $P$ is visible from some points in $H_v$, any points in the sinks of $P$ is also visible from some points in $H_v$. It contradicts the hypothesis that $G_v$ is the minimum set that covers all VG-sinks, completing the forward proposition. The converse is trivial.
\end{pf}

\noindent \textbf{Algorithm}\\
Step 1. Find visibility polygons $P_i=VP(P,v_i)$ for $\forall v_i\in V(P)$ where V(P) is the set of all vertices of polygon $P$.\\
Step 2. Compute all visibility regions $R_1,R_2,\dots,R_r$ by constructing the planar subdivision of $W=\cup _{1\leq i\leq n}(\partial P_i-\partial P)=\cup _{1\leq i\leq n}W_i$  where $\partial P_i$ is boundary of visibility polygon $P_i$.\\
Step 3. Construct a dual graph $D$ of visibility regions.\\
Step 4. Find the set of sinks $S=\{s_1,s_2,\dots,s_m\}$ of $G$.\\
Step 5. Compute the set of visible vertices $VV_j$ for each sink $s_j\in S$.\\
Step 6. Compute the set of visible sinks $U_i$ for each vertex $v_i\in V(P)$.\\
Step 7. Solve the set-covering problem of $U_1,U_2,¡¦.,U_n$.\\
\null

\noindent \textbf{Analysis}
Let us analyze the time complexity of Step 1-4. Since $VP(P,v_i)$ can be computed by linear algorithm \cite{Gin81,Joe87,Lee83}, Step 1 requires $O(n^2)$ time. The number of intersection pairs in $W$ is $O(n^3)$ \cite{Bos91,Bos02} and $|W|=O(n^2)$. Thus, the planar subdivision of $W$ can be constructed in $O(n^2\log n^2+n^3)=O(n^3)$ time \cite{Bal95,Cha92}. By the theorem in \cite{Bos91,Bos02} that the number of visibility regions $r\leq O(n^3)$, Step 3 can be done in $O(n^3)$ by considering all the edges of visibility regions from the segments in $W$. Step 4 can be done easily in $O(n^3)$ by searching all the nodes of $D$. By Lemma \ref{lemma1}, it is enough to compute the minimum set $G_v$ that covers $S$. $G_v$ can be computed in Step 5-7. In Step 5, since each $s_j$ is a convex region \cite{Bos91,Bos02}, a certain point $p_j$ in $s_j$ can be easily computed. Then $VV_j=\{v |v\in VP(P,p_j),v\in V(P)\}$ can be computed in $O(n)$ time using any linear time visibility polygon algorithm \cite{Gin81,Joe87,Lee83}. Since $m\leq O(n^2)$ \cite{Bos91,Bos02}, Step 5 takes $O(mn)=O(n^3)$ time. Step 6 can be done easily in $O(n^3)$ times by considering all pairs of $s_j\in S$ and $v_i\in V(P)$. Step 7 can be done in $O(mn)=O(n^3)$ time with the approximation ratio of $O(\log n)$ \cite{Chv79}.

\begin{thm}
An approximate solution that is at most $O(\log{n})$ times the optimal solution can be computed by the approximation algorithm of $O(n^3)$ time complexity for VG problem in any simple polygon.
\end{thm}

\begin{figure}
\centerline{
    \includegraphics[height=2in]{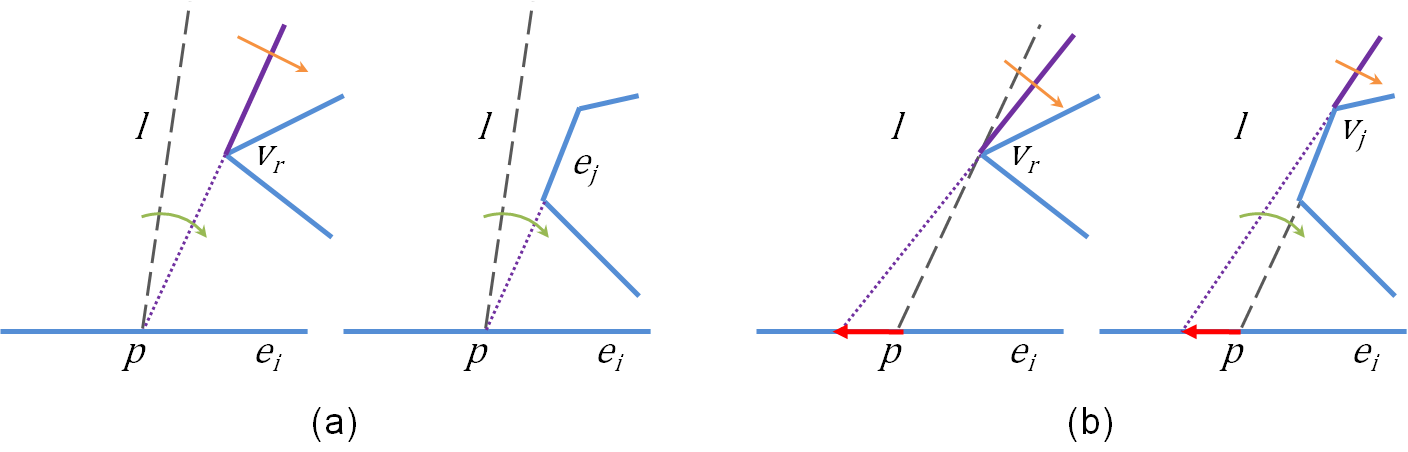}}
    \caption{Available rotations of line of sights from points on an edge $e_i$\label{fig2}}
\end{figure}

\section{Approximation Algorithm for Edge Guards}

\begin{figure}
\centerline{
    \includegraphics[height=2in]{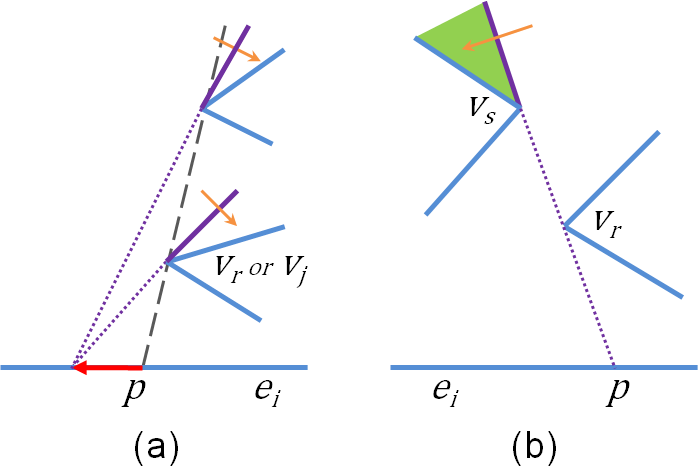}}
    \caption{(a) A series of reflex vertices on the same side can not block the further rotation of line of sight $l$ (b) The case that a pair of reflex vertices stop the CCW rotation of $l$ and make a window of $p$ on $e$\label{fig3}}
\end{figure}

\begin{lem}\label{lemma2}
The set of all windows for EG is a subset of the set of all windows for VG in simple polygon and has the same orientation in the dual graph.
\end{lem}
\begin{pf}
We can imagine a line of sight $l$ from a point $p$ on $e_i$. Let $p$ denote a \emph{source} of $l$. $l$ can be rotated until it meet a reflex vertex $v_r$ of $P$ or one of other edges $e_j$ becomes collinear to it (Fig.\ref{fig2}.(a)). After that, it can still rotate using other points along $e$ as a source and $v_r$ or $v_j$ as a pivot (Fig.\ref{fig2}.(b)). Other successive reflex vertices from the same side of $v_r$ or $v_j$ can not obstruct the rotation (Fig.\ref{fig3}.(a)). This rotation can be stopped at the end of $e_i$, i.e. a vertex of $P$, thus the windows from this situation is identical to the windows from VG problem. Fig.\ref{fig3}.(b) shows a window for the edge $e$ that does not use any vertex on $e$ as its source and it is the only case that can occurs in a simple polygon. $l$ can not rotate in CCW any more and any point in shaded(green colored) area is not visible from any point on $e_i$. This window is exactly the same as the window of $v_r$ using $v_s$ as a base thus orientation of corresponding edges in the dual graph is also same. The case of opposite direction can be done in the same way. Therefore, any window from the source on an edges of $P$ is an element of the set of all windows for VG and has the same orientation in the dual graph.
\end{pf}

\begin{lem}\label{lemma3}
The set of EGs $G_e$ is the optimal solution of EG problem in a simple polygon $P$ if and only if $G_e$ is the minimum set that covers all VG-sinks.
\end{lem}
\begin{pf}
Suppose a sink $s$ of visibility regions constructed by the set of all windows for EG is weakly visible from an edge $e$. By Lemma \ref{lemma2}, the windows of $e$ are the same as some windows in the set of all windows for VG and has the same orientation in the dual graph. Since $s$ is weakly visible from $e$, it should be in the set of points in $P$ that are weakly visible from $e$. Let this set denote visibility polygon $VP(P,e)$ of $e$. $VP(P,e)$ is bounded by the some parts of boundary of $P$ and windows of $e$ that have edges in dual graph with outward directions from $VP(P,e)$. Therefore, $s$ can not be reachable from any visibility region in the outside of $VP(P,e)$ and all regions in the span $K$ of $s$ are in the $VP(P,e)$; all regions in the span $K$ is weakly visible from $e$. By the same logic in the proof of Lemma \ref{lemma1}, Lemma \ref{lemma3} is concluded.

\end{pf}

For the following description of an algorithm for EG problem, we define two more terminologies. $E(P)$ is the set of all edges on the boundary of $P$. An edge $e\in E(P)$ is partially visible from a convex region $R\subseteq P$ if all the points in $R$ are weakly visible from $e$.\\

\noindent \textbf{Algorithm}\\
Step 1. Find visibility polygons $P_i=VP(P,v_i)$ for $\forall v_i\in V(P)$.\\
Step 2. Compute all visibility regions $R_1,R_2,\dots,R_r$ by constructing the planar subdivision of $W=\cup _{1\leq i\leq n}(\partial P_i-\partial P)=\cup _{1\leq i\leq n}W_i$.\\
Step 3. Construct a dual graph $D$ of visibility regions.\\
Step 4. Find the set of sinks $S=\{s_1,s_2,\dots,s_m\}$ of $G$.\\
Step 5. Compute the set of partially visible edges $VE_j$ from each sink $s_j\in S$.\\
Step 6. Compute the set of weakly visible sinks $U_k$ from each edges $e_k\in E(P)$.\\
Step 7. Solve the set-covering problem of $U_1,U_2,¡¦.,U_n$.\\
\null

\noindent \textbf{Analysis}
We can compute $S$ in $O(n^3)$ time from the analysis of algorithm for VG problem. By Lemma \ref{lemma3}, it is enough to compute the minimum set $G_e$ that covers $S$. $G_e$ can be computed in Step 5-7. In Step 5, since each $s_j$ is a convex region \cite{Bos91,Bos02}, a certain point $p_j$ in $s_j$ can be easily computed. Then, $VE_j=\{e|\exists_{x\in e}(x\in VP(P,p_j)),e\in E(P)\}$ can be computed in $O(n)$ time using any linear time visibility polygon algorithm \cite{Gin81,Joe87,Lee83}. Since $m\leq O(n^2)$ \cite{Bos91,Bos02}, Step 5 takes $O(mn)=O(n^3)$ time. Step 6 can be done easily in $O(n^3)$ times by considering all pairs of $s_j\in S$ and $e_k\in E(P)$. Step 7 can be done in $O(mn)=O(n^3)$ time with the approximation ratio of $O(\log n)$ \cite{Chv79}.

\begin{thm}
An approximate solution that is at most $O(\log{n})$ times the optimal solution can be computed by the approximation algorithm of $O(n^3)$ time complexity for EG problem in any simple polygon.
\end{thm}

\section{Concluding Remarks}
For any simple polygon $P$ of $n$ vertices, an approximation solution of VG, EG problem can be computed in $O(n^3)$ time and the size of the solution is at most $O(\log{n})$ times the optimal. As in \cite{Gho10}, presented algorithms can be adopted for the VG and EG problems in polygons with holes. But, the number of sinks are $O(n^4)$ that is the same complexity of all visibility regions in polygons with holes \cite{Zar05}. Therefore, no further reduction on time complexity than $O(n^5)$ algorithms in \cite{Gho10} can be archived unless there exist some smaller dominant sets as sinks in simple polygons.

\bibliographystyle{abbrv}
\bibliography{my_article}

\nocite{Agg88,Avi81,Bal95,Bos91,Bos02,Cha92,Chv79,Chv75,Ede84,Efr06,Gho10,Gho87,Gin81,Joe87,kah83,Lee86,Lee83,ORo83-2,ORo83-1,She92,Zar05}

\end{document}